# Monte Carlo simulation of charge transport in disordered organic systems using buffer lattice at boundary

S. Raj Mohan[1*], Manoranjan P. Singh[2,3], M. P. Joshi[1,3]

[1]Laser Materials Processing Division, Raja Ramanna Centre for Advanced Technology, Indore, India, 452013.

[2]Theory and Simulation Lab, Raja Ramanna Centre for Advanced Technology, Indore, India, 452 013.

[3]Homi Bhabha National Institute, Training School Complex, Anushakti Nagar, Mumbai, India, 400094.

**Abstract**

In this article, we present an alternative method for simulating charge transport in disordered organic materials by using a buffer lattice at the boundary. This method does not require careful tracking of carrier's hopping pattern across boundaries. Suitability of this method is established by reproducing the field dependence of mobility, carrier relaxation and carrier diffusion in disordered organic systems obtained by simulating the charge transport for the full length of the systems along the field direction without any boundary condition. The significance of the buffer lattice is emphasized by simulating field dependence of mobility without using a buffer lattice, which results in negative field dependence of mobility (NFDM) at low field regime due to the extra bias the carrier gains from the neglected hops at boundaries along field direction.

---

[*]Corresponding author. Tel.: +91 7312488304; fax: +91 7312488300
E-mail address: raj@rrcat.gov.in; rajmho@gmail.com (S. Raj Mohan).

# 1 Introduction

Gaussian disorder model (GDM) [1-4] is a widely used model for explaining the charge transport behavior observed in disordered organic systems. According to GDM the charge transport in disordered organic materials occur by hopping among transport sites that are subjected to energetic and positional disorder [1-4]. Since the model assumes Gaussian density of states [1-4] a complete analytical solution of the hopping transport is therefore difficult, especially in 3-D. Hence, the predictions of the GDM were made on the basis of Monte Carlo simulation of hopping charge transport [1-4]. Monte Carlo simulation is considered as an idealized experiment with which one can study the charge transport in disorder system as function of several of parameters[1-5]. Generally, charge transport is simulated for thin films with thickness of few microns. This is to make sure that the carrier has attained a dynamic equilibrium during the transit and also to have a better comparison with the experiments, like Time of flight (TOF), which are generally performed on micron size thick samples [1-4]. Periodic boundary condition (PBC) is a well established and frequently employed [6-11] to simulate the charge transport in micron thick thin films. PBC is a set of boundary conditions that are used to simulate the properties of bulk system by simulating a part of it [12]. In principle, PBC generates an infinitely large system with help of a smaller array, that represent only a part of the bulk system, with the assumption that the small array will replicate periodically in all the three directions to form the bulk system. In PBC, when a carrier moves out through one of the boundaries a similar carrier is injected in through the opposite boundary [12]. In this process the carrier's energy and the other Cartesian coordinates, other than the directions along which the PBC is applied, remain same as at the boundary. In essence, the advantage of using the PBC is that the simulation can be performed on a sample length of several microns using an array of smaller size while at the same time reproducing the results obtained when simulated without using PBC. If PBC is not employed then an array of bigger sizes, in the all the three Cartesian directions, is required which demands large computational resources.

A meticulous implementation of carrier's hopping pattern across boundaries is indispensable for implementing PBC. All backward and forward hops across the boundaries must be taken care of otherwise serious artifacts may arise. In this work, we propose an alternative method of implementing the simulation with the help of buffer lattice region at the boundaries. Buffer regions have been commonly employed for simulations in other areas of research[13] but the use of buffer lattice in Monte Carlo simulation of charge transport in disordered organic system [1-4] has not been reported before. In this method, when the carrier reaches at a defined final boundary plane in the field direction then the carrier is taken to another defined plane of the lattice in the opposite direction, where the carrier gets a buffer lattice region. In the buffer lattice region carrier is allowed to perform all the hopping process that it would have made in the absence of a boundary. This method does not require the stringent tracking of carrier's hopping pattern. Accuracy of this method is established by reproducing the field dependence of mobility in disordered organic systems obtained by simulating the charge transport for the full length of the sample along the field direction. In order to emphasize the role of buffer lattice the field dependence of mobility is simulated without using the buffer lattice along the field direction and the data is compared with the one obtained by FL simulation (FLS). The discrepancy is observed mainly at the low field regime. At low field regime, field dependence of mobility when simulated without buffer lattice and with zero positional disorder shows negative field dependence of mobility (NFDM) for all values of energetic disorder. While by FLS a clear saturation of mobility with field is observed at low field regime. In the absence of buffer lattice some of the hops that the carrier may make in the absence of such a boundary are neglected. These neglected hops give an extra bias to the carrier which results in the enhanced mobility that lead to NFDM. Thus, this study also highlights the importance of the flawless implementation of the carrier's hopping pattern at boundaries while implementing PBC, failing which results in a serious artifact that can mislead the interpretation and the modeling of charge transport in disordered organic systems.

## 2 Details of Monte Carlo simulation

The Monte Carlo simulation is based on the commonly used algorithm reported by Schönherr et al [1,14]. A 3D array is considered as the lattice with size 70x70 along *x* and *y* direction. Along *z* direction, the direction of the applied field, various sizes are used to implement different simulation approaches adopted for covering the required sample thickness along this direction. Simulation is performed for a sample length of 4μm along field direction. The lattice constant $a = 6$Å is used for the entire simulation [1]. Three different approaches are adopted:

***Case 1 (FLS):*** In this case a lattice of size 70x70x7000 along *x, y and z* direction is used for simulation. Simulation is performed without using PBC along *z* direction (Fig.1(a)), which requires array of bigger size. However, PBC is implemented along *x* and *y* direction. In this case, carrier does not see any boundary along *z* direction till it covers the required sample thickness.

***Case 2:*** This is an alternative method for simulating the charge transport (Fig.1(b)) by using buffer lattice at boundaries. In this case a lattice of size 70x70x150 along *x, y* and *z* direction is used. Carrier injected into *z*=1 plane is allowed to move in the direction of the applied electric field.

The carrier is taken into $z=70^{th}$ plane once it crosses $z=140^{th}$ plane while keeping the *x* and *y* coordinates and the carrier energy remain same. Therefore, the lattice region up to $z=69$ acts as buffer lattice for the carrier taken to the $70^{th}$ plane. Around $z=70^{th}$ plane, the carrier performs all the hops that it would have made in the absence of a boundary. Similarly, the carrier reaches any boundary along *x* and *y* direction is taken to the middle of the same plane (keeping the *z* coordinate same) with same energy. This approach considers all the carrier hops at the boundaries perpendicular to *z* direction because carrier is allowed to execute the necessary hops inside the lattice. Use of buffer lattice region along *x* and *y* directions and proceed as in the case of *z* direction is also possible. But, this requires bigger array for providing buffer lattices around the boundaries along *x* and *y* directions. In general, this method is an alternative method for simulating the charge transport in disordered organic systems at a reasonable computational resource. *Case 2* requires less computational resources compared to *case 1*. More over, this method doesn't require the careful tracking of the carrier's hopping pattern.

*Case 3:* In this case a lattice of size 70x70x70 along *x, y and z* direction is used for simulation (Fig.1(c)). This method of simulation and analysis of the data emphasizes and illustrates the role of buffer lattice adopted in *case 2* described above. In this case, PBC is implemented only along x and y direction. Along the applied field direction the carrier is taken to first plane ($z=1$) once the carrier reaches $70^{th}$ plane, keeping the carrier energy and the other coordinates same as at the boundary. In this method, the carrier taken to the first plane cannot perform backward hop at the boundary along the field direction, i.e. carrier taken to the first plane proceed in the field direction like a carrier injected initially but with the relaxed energy it has attained during the hopping process. Hence, some of the hops that the carrier may make in the absence of boundaries, which it encounters in the process of transit along *z* directions, are neglected. In *case 2*, carrier performs these neglected hops in the buffer region. Hence, the data obtained by this method not only emphasizes the role of buffer lattice in *case 2* but also highlights the possible artifact that can arise if carrier's hopping pattern across boundaries is not properly taken care of while implementing PBC.

In each of these cases mentioned above, the hopping of carrier between sites *i* and *j* is decided by the probability that a carrier jumps from the present site *i* to another site *j* around and within a cube of size 7x7x7 (343 sites)[10]. Therefore, in this *case 2*, the buffer lattice between $z=140$ and $z=150$ eases the probability calculation for carrier hopping (within 7x7x7 sites around the carrier) at the boundary, instead of dealing with the sites at the opposite boundaries. Each node of the array is considered as a localized transport site with uncorrelated site energy. The uncorrelated site energies of the lattice are taken randomly from a Gaussian distribution with a known standard deviation ($\sigma$), which is the measure of the energetic disorder of the sample. Simulations are performed for various value of $\sigma$ and electric field strengths at 300K. Throughout the simulation the positional disorder is neglected ($\Sigma=0$). This is to avoid the huge computational time required for simulating the charge transport with non-zero positional disorder. Moreover, the outcome of this study can be clearly shown even without zero positional disorder. Carrier hopping in this energetically disordered lattice is governed by Miller-Abrahams equation [15]. Miller-Abrahams equation for the jump rate ($v_{ij}$) of the charge carrier from the site $i$ to site $j$ is given by,

$$v_{ij} = v_0 \exp\left(-2\gamma a \frac{\Delta R_{ij}}{a}\right) \begin{pmatrix} \exp\left(-\frac{\varepsilon_j' - \varepsilon_i'}{kT}\right) &,& \varepsilon_j' > \varepsilon_i' \\ \\ 1 &,& \varepsilon_i' > \varepsilon_j' \end{pmatrix} \qquad (1)$$

where, $\varepsilon_i'$ and $\varepsilon_j'$ are the effective energies of the site $i$ and $j$ including the electrostatic energy, $a$ is the intersite distance, $\Delta R_{ij} = |R_i - R_j|$ is the distance between sites $i$ and $j$, $k$ is the Boltzmann constant, $T$ is the temperature in Kelvin and $2\gamma a$ is the wave function overlap parameter which controls the electronic exchange interaction between sites. Throughout the simulation we took $2\gamma a = 10$ [1,14]. Transit time of a carrier is calculated by adding all the hopping times and averaging over hundred carriers. The mobility is calculated using drift mobility equation. The electric field range (>$10^4$ V/cm) over which the simulations are carried out is higher and hence, the diffusion cannot dominate the charge transport [16,17]. Thus, the use of drift mobility equation is justified.

## 3 Results and discussions

Fig. 2 shows the comparison of the field dependence of mobility, for various values of energetic disorder, obtained from *case 1* and *case 2*. For both the cases, field dependence of mobility for all the values of energetic disorder under study is similar and they super impose on each other. Inset shows that the total number of hops that the carriers takes for traversing the required sample length. There is not remarkable difference in total number of hops made by the carrier in traversing the sample when simulated using *case 1* and *case 2*. In addition to the field dependence of mobility other data such as relaxation, diffusion etc calculated using *case 2* are also identical to that obtained using *case 1*. Figure 3(a) and 3(b) shows the temporal changes in the mean energy and diffusion of carriers during the transit. Both the energy of the carriers and the diffusion has attained a steady state value which confirms that in both cases carrier has attained dynamic equilibrium [4,18]. In both cases, the temporal variation in the carrier mean energy and diffusion occurs in similar manner and attained their respective steady state values in the same time frame. This clearly shows for the that the charge transport in disordered organic system can be accurately simulated also by using buffer lattice at boundaries as in *case 2*. Inside the buffer lattice carrier makes all the necessary hops that it would have made in the absence of such a boundary. *Case 2* takes less computational space requirement than *case 1*. A precise following of carrier's hopping pattern is also not necessary when simulation is carried out as in *case 2*.

In order to emphasize the role of buffer lattice in *case 2*, the field dependence of mobility is also simulated without using buffer lattice (*case 3*) and compared with the data obtained using *case 1*. Field dependence of mobility, for various values of energetic disorder, obtained from *case 1* and *case 3* is shown in Fig. 4. For both the cases, field dependence of mobility for all the values of energetic disorder under study is similar except at lower electric field strengths (~<$3.6 \times 10^5$ V/cm). In *case 3*, at lower electric field strengths, the mobility first decreases with increase of electric field, ie. NFDM. Mobility attains a minimum value (indicated by arrows in Fig. 4) and then shows a positive field dependence given by $ln\mu \propto E^{1/2}$ as predicted by GDM [1-4]. In *case 1*, for all the values of energetic disorder, a clear saturation of mobility is observed at lower electric field strengths [1-4]. NFDM is observed in *case 3* without any positional disorder ($\Sigma=0$). The strength of observed NFDM becomes remarkable when the energetic disorder decreases, as shown in Fig. 5. The strength of NFDM ($\Delta\mu_{NFDM}$) is assigned as the difference in mobility value for the lowest electric field strength under study and the observed minima of the mobility at low field regime. The origin of the difference in field dependence of mobility, at lower electric field strengths, between *case 1* and *case 2* is explained below.

Fig. 5 shows the variation of the total number of hops with electric field, for various values of energetic disorder, executed by the carrier to cover the required sample length when simulated with *case 1* and *case 3*. At low electric field strengths, for all the values of energetic disorder, the total number of hops made by the carrier to cover the required sample length in *case 1* is higher than in *case 3*. Maximum value of $\Delta hops$ occurs at lowest electric field strength ($4 \times 10^4$ V/cm), which decreases as the electric field strength increases and become negligible at very high electric field strengths. As the value of energetic disorder increases the value of *Δhops* also increases (inset of Fig. 5). The neglected hops, in *case 3*, certainly influences the charge transport especially the transit time. But, it is confirmed through simulation that the charge carriers have attained dynamic equilibrium [4,18] while covering the required sample length. The neglected hops provides an extra bias for the carrier to move in the applied field direction. Carrier covers the required sample length in less number of hops, which reduces the carrier transit time and thereby enhances the mobility. The maximum difference in transit time ($\Delta\tau$) between *case 1* and *case 3* occurs at the highest value of *Δhops*. Thus, the strength of the NFDM in *case 3* is higher at lower electric field strength for any value of energetic disorder. As the electric field strength increases *Δhops* and the strength of NFDM diminishes concurrently. Similarly, for a constant electric field, the strength of the observed NFDM in *case 3* is expected to be higher at higher value of energetic disorder. But, higher NFDM is observed at lower energetic disorder (Figure 4). This is because the difference in mobility, for a constant electric field and thickness, depends on the value of $\xi$ instead on $\Delta\tau$ alone, as shown in equation (2).

$$\Delta\mu = \mu_2 - \mu_1 = \frac{L}{E}\left[\frac{1}{\tau_2} - \frac{1}{\tau_1}\right] = \frac{L}{E}\xi \qquad (2)$$

Where, $L$ is the thickness of sample, $E$ is the applied electric field, $\xi = \left[\frac{\Delta\tau}{\tau_2\tau_1}\right]$, $\Delta\tau$ is the difference in transit time between *case 1* and *case 2*, $\tau_2$ is transit time in *case 3* and $\tau_1$ is the transit time in *case 1*. Fig. 6a and 6b shows the variation of $\xi$ as a function of electric field and energetic disorder. Electric field is limited to the range where the value of $\Delta\tau$ is significant. For any value of energetic disorder, $\xi$ decreases with increase of electric field strengths (Fig. 6(a)). This supports the observed NFDM at low field regime for all values of energetic disorder. In addition, $\xi$ decreases with increase of energetic disorder (Fig. 6(b)). This suggests a strong NFDM for low value of energetic disorder and vice versa, as shown in Fig. 4. The above data clearly shows an artifact that occurs at low electric field regime due to hops neglected when simulated using *case 3*. Thus, the data obtained from *case 3* not only highlights the role of buffer lattice used in *case 2* but also asserts the need for precise consideration of carrier hopping pattern across boundaries while implementing PBC for simulating charge transport in disordered organic materials.

Therefore, in *case 2*, enough buffer array must be provided so that carrier should not reach the $z=1$ plane after it has taken to $z=70^{th}$ plane. After taking in to $z=70^{th}$ plane, if the carrier hops back and touches $z=1$ plane then some hops may be neglected which in turn may result in NFDM. Hence, a sufficient buffer lattice must be provided for the wandering of the carrier around the $z=70^{th}$ plane. The dimension of the buffer lattice used in the study (69x69x69) is optimum for simulating field dependence of mobility and other parameters for various energetic disorders under study.

**4 Conclusion**

In this study, a simulation method that excludes the use of PBC in Monte Carlo simulation of charge transport in disordered organic systems is demonstrated. This method adopts the use of buffer lattices at the boundaries for the simulation. In this method a carrier reaching at one boundary plane is taken to another plane, where the carrier is provided with a buffer lattice region. Inside the buffer lattice region the carrier makes all the necessary hops that it would have made in the absence of a boundary. The importance of buffer lattice is emphasized by simulating the field dependence of mobility without using a buffer lattice which results in NFDM at low field regime. The observed NFDM is due to the extra bias the carrier gains from the neglected hops at boundaries in the absence of buffer lattice region.

# Figure Captions

Figure 1. Schematic diagram showing a typical hopping motion of carrier inside the lattice for *case 1* (a), *case 2* (b) *case 3* (c). Shaded region in (a) shows the hops that are neglected at boundaries upon implementing the *case 3*.

Figure 2. Comparison of the field dependence of mobility simulated for (●) *case 1* and (○) *case 2*. Inset shows the comparison of total number of hops made by carrier in (▲) *case 1* and (○) *case 2*.

Figure 3. Comparison of the relaxation (a) and diffusion (b) of carriers simulated using *case 1* and *case 2*. Open symbols shows data obtained using *case 2*. Simulation is carried for E=7.5x10$^4$V/cm and T=300K.

Figure 4. Comparison of field dependence of mobility for various values of disorder for (○) *case 3* and (●) *case 1*. Arrow shows the minima of mobility occurred at low field regime in *case 3*. Enlarged view of low field regime for the respective cases is shown in the inset.

Figure 5. Variation of the strength of NFDM, observed in *case 3*, with energetic disorder. Solid line is a guide to eye.

Figure 6. Field dependence of the total number of hops made by the carrier in two cases (*case 3* (solid line) and *case 1* (dashed line)), for various values of energetic disorder. Inset shows the dependence of difference in total number of hops (*Δhops*), made by the carrier in *case 1* and *case 3*, with energetic disorder at 4x10$^4$V/cm.

Figure 7. Variation of $\xi$ (a) with electric field strength for the values energetic disorder where reasonably strong NFDM is observed (b) with energetic disorder for the various value of electric fields at low field regime. Solid line is a guide to eye.

**Figure 1**

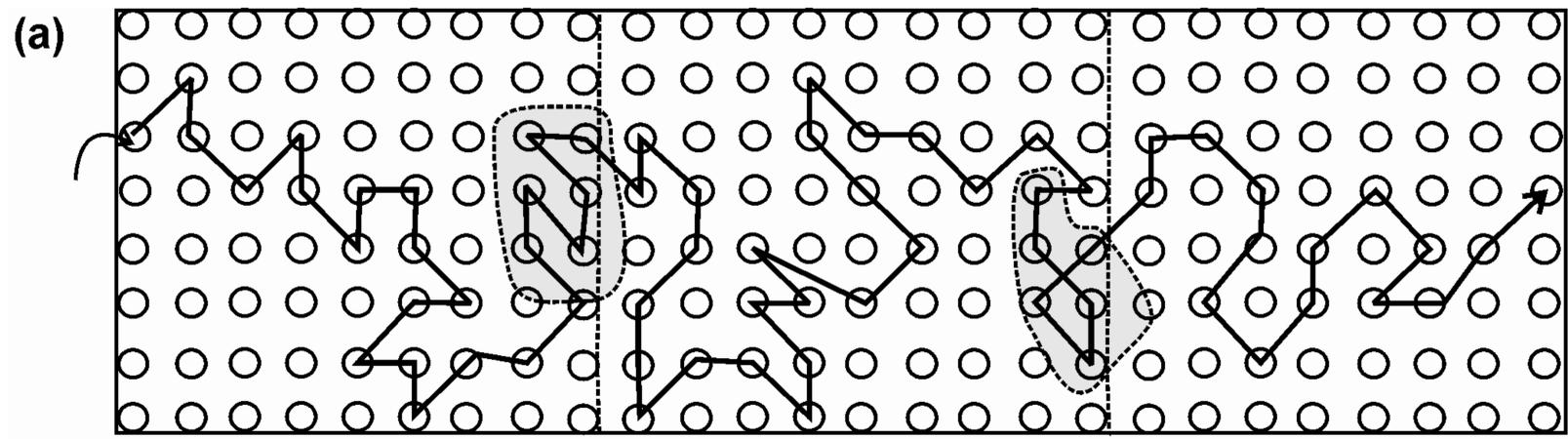

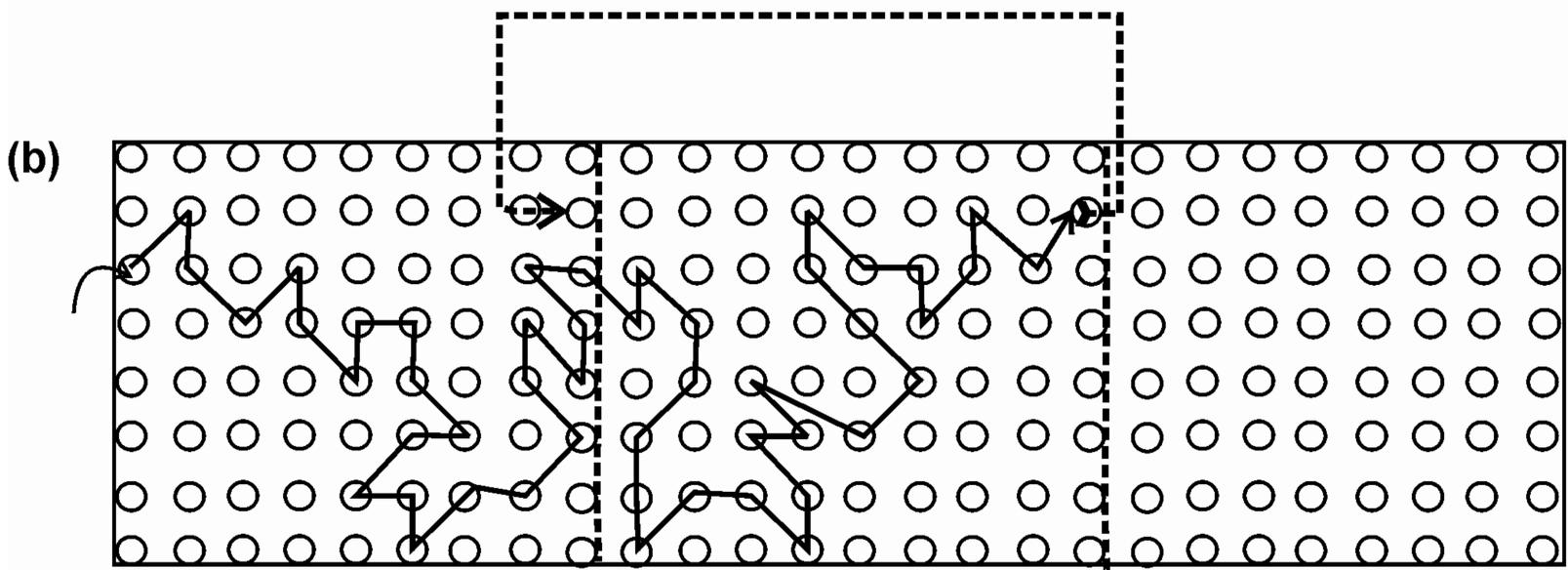

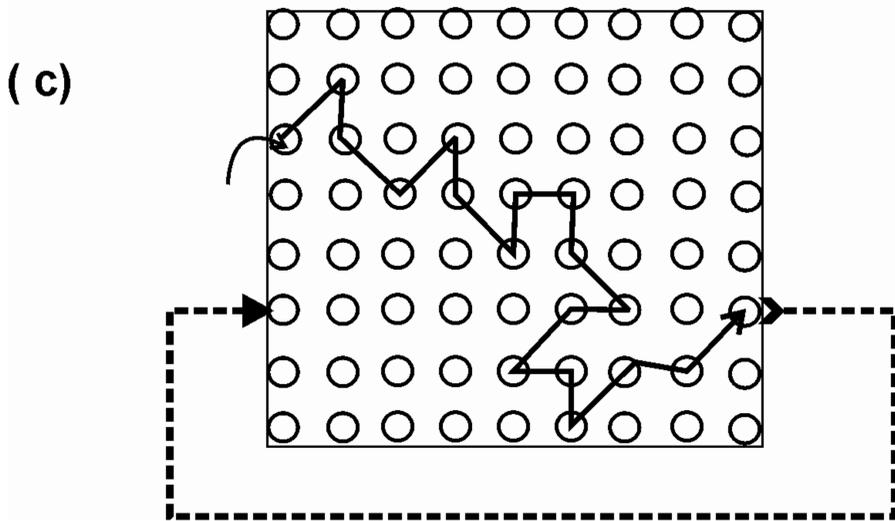

**Figure 2**

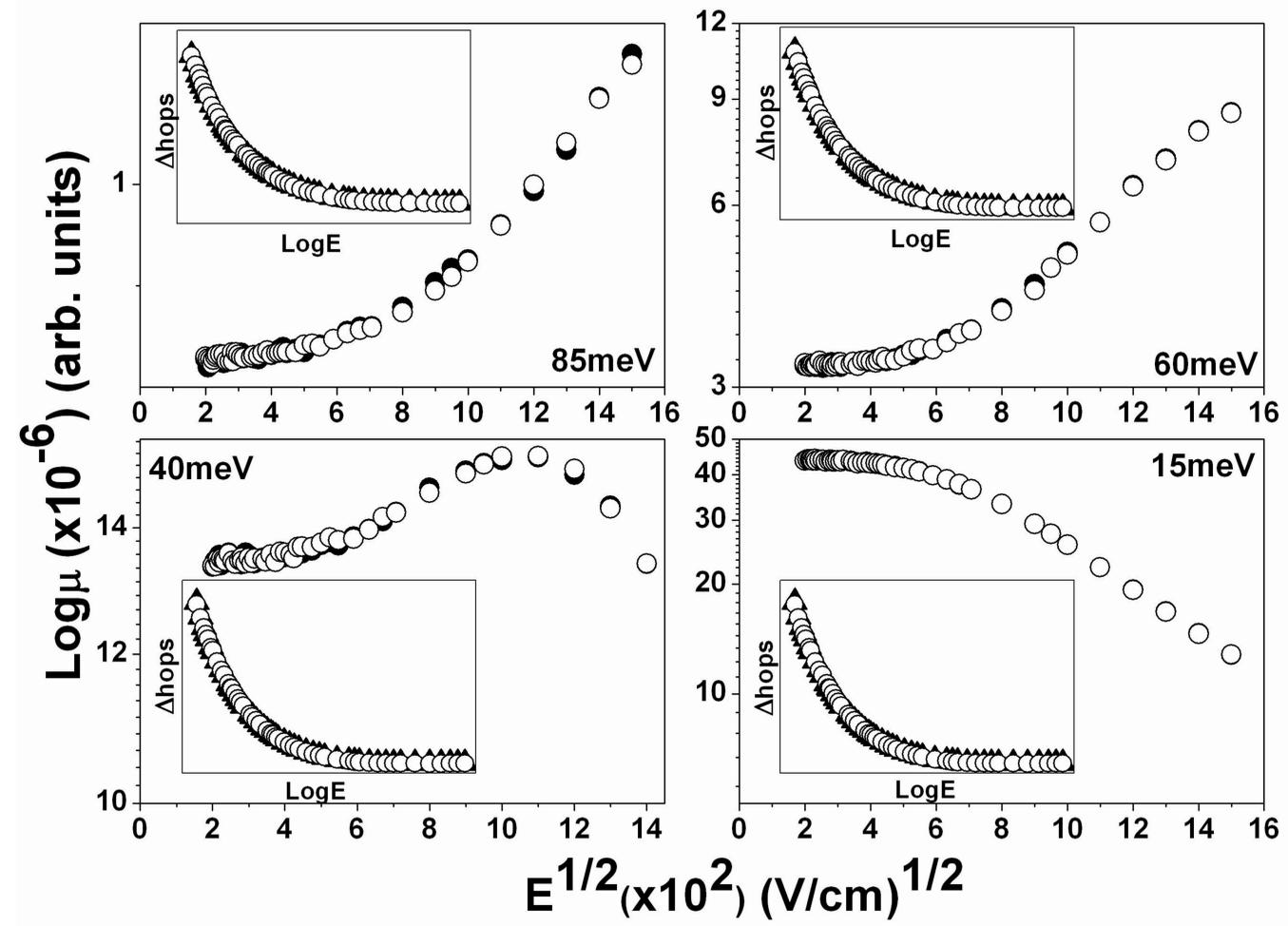

Figure 3

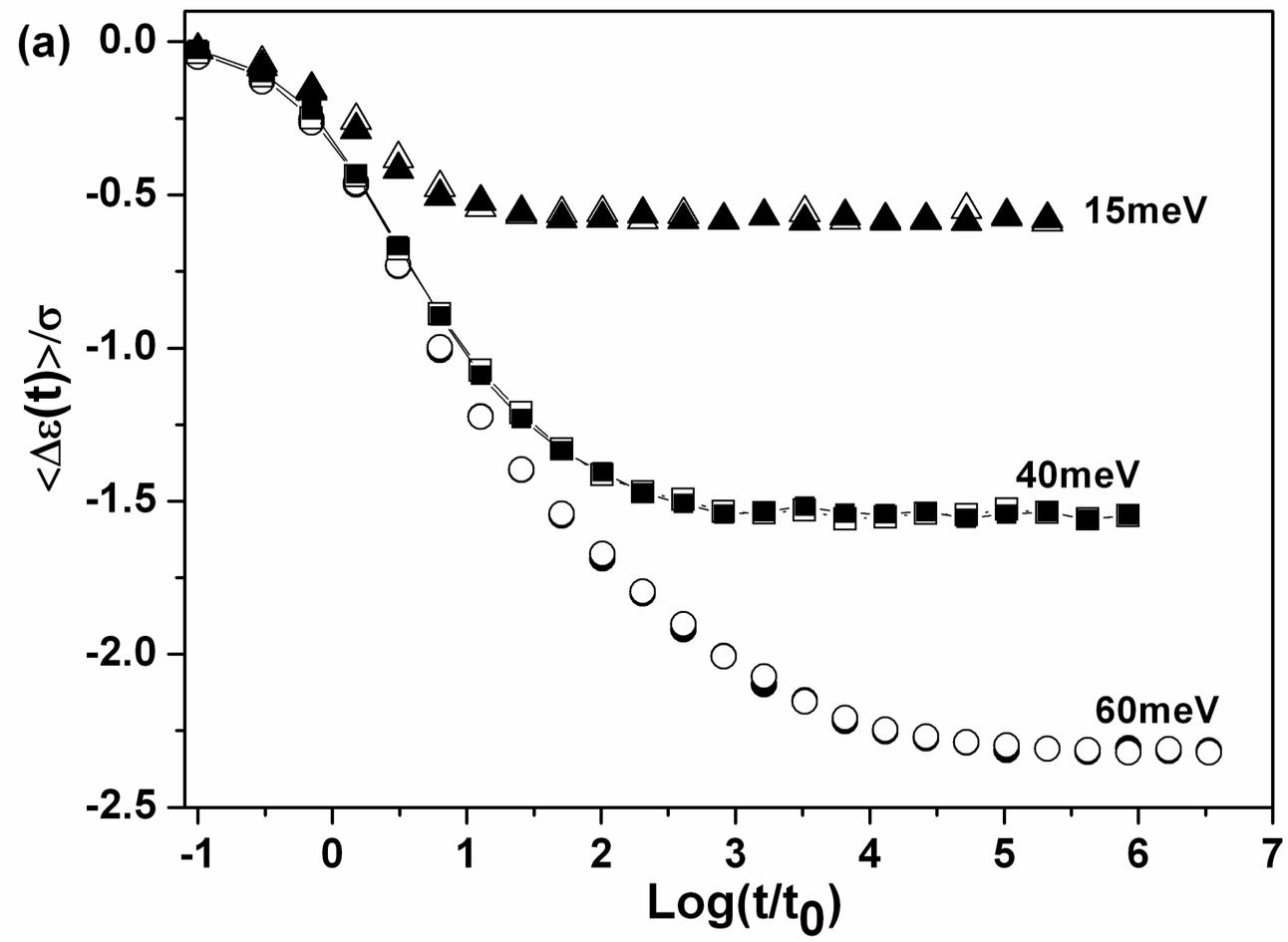

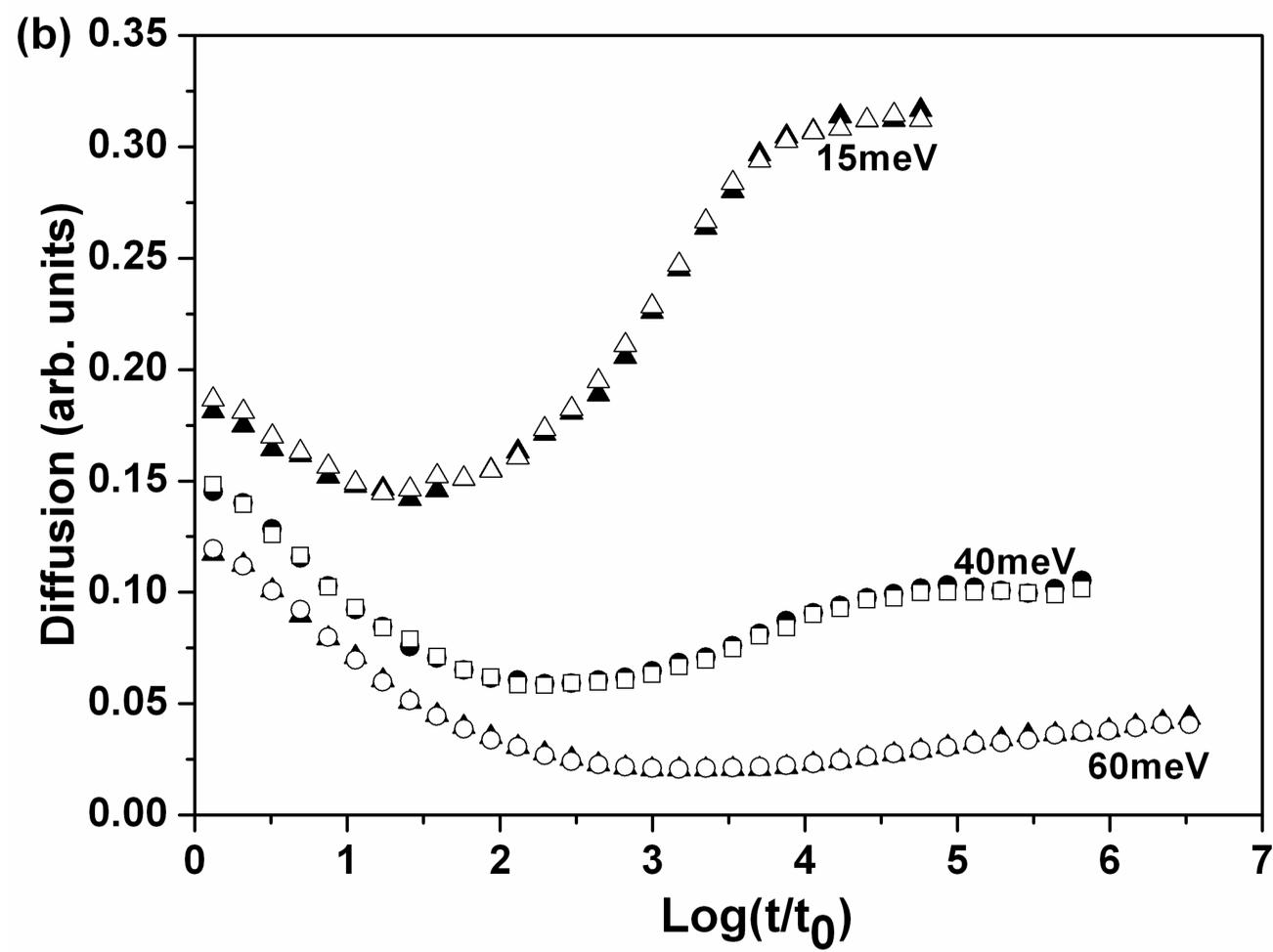

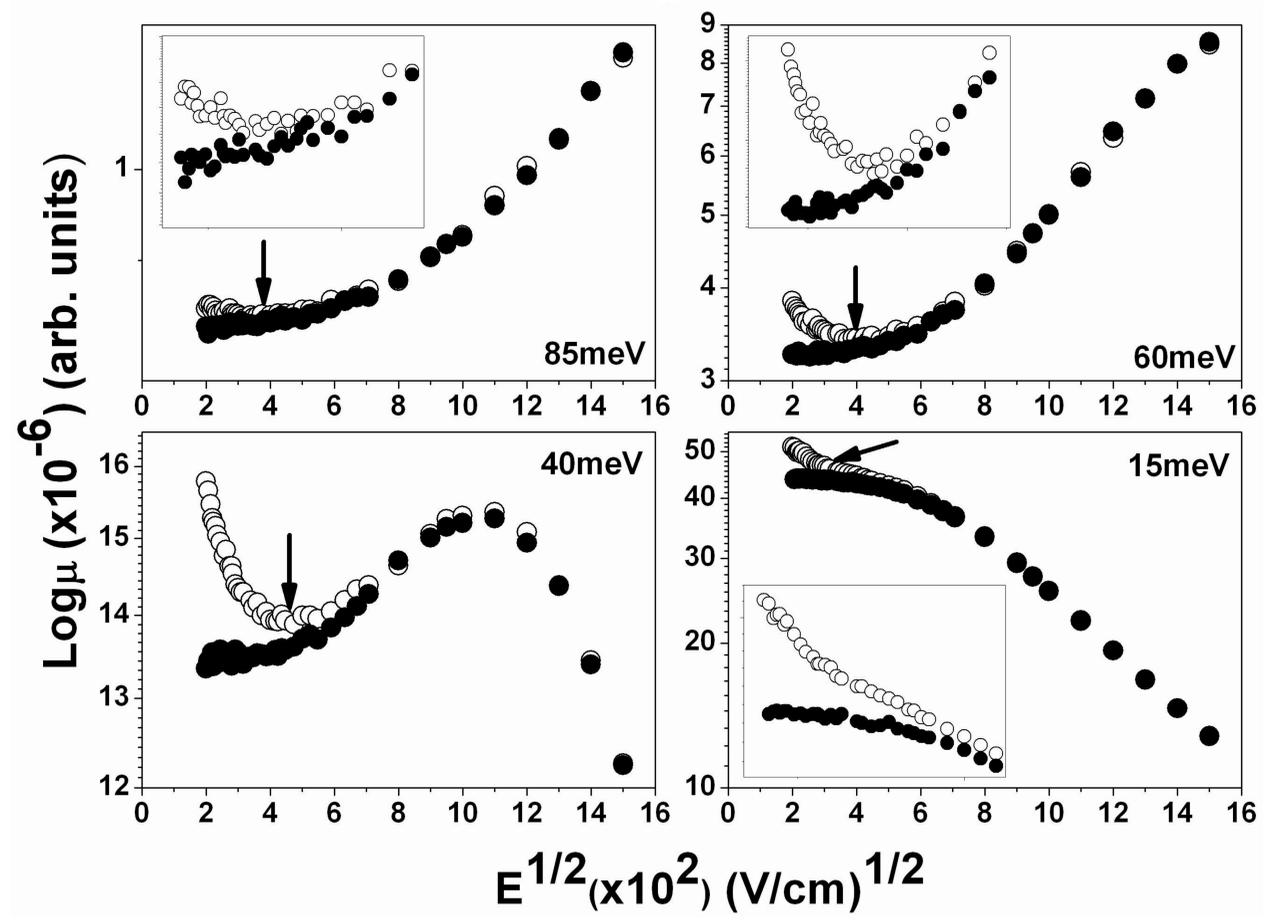

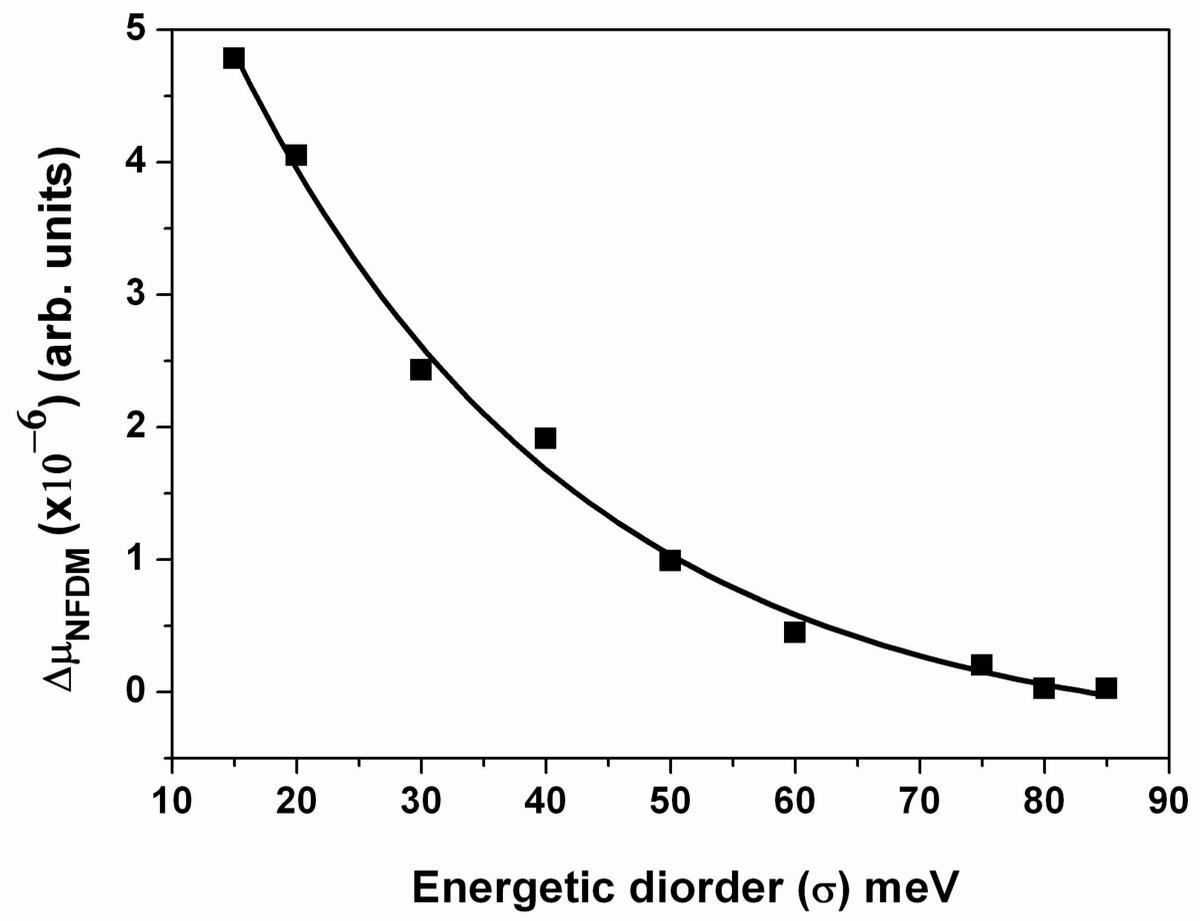

**Figure 6**

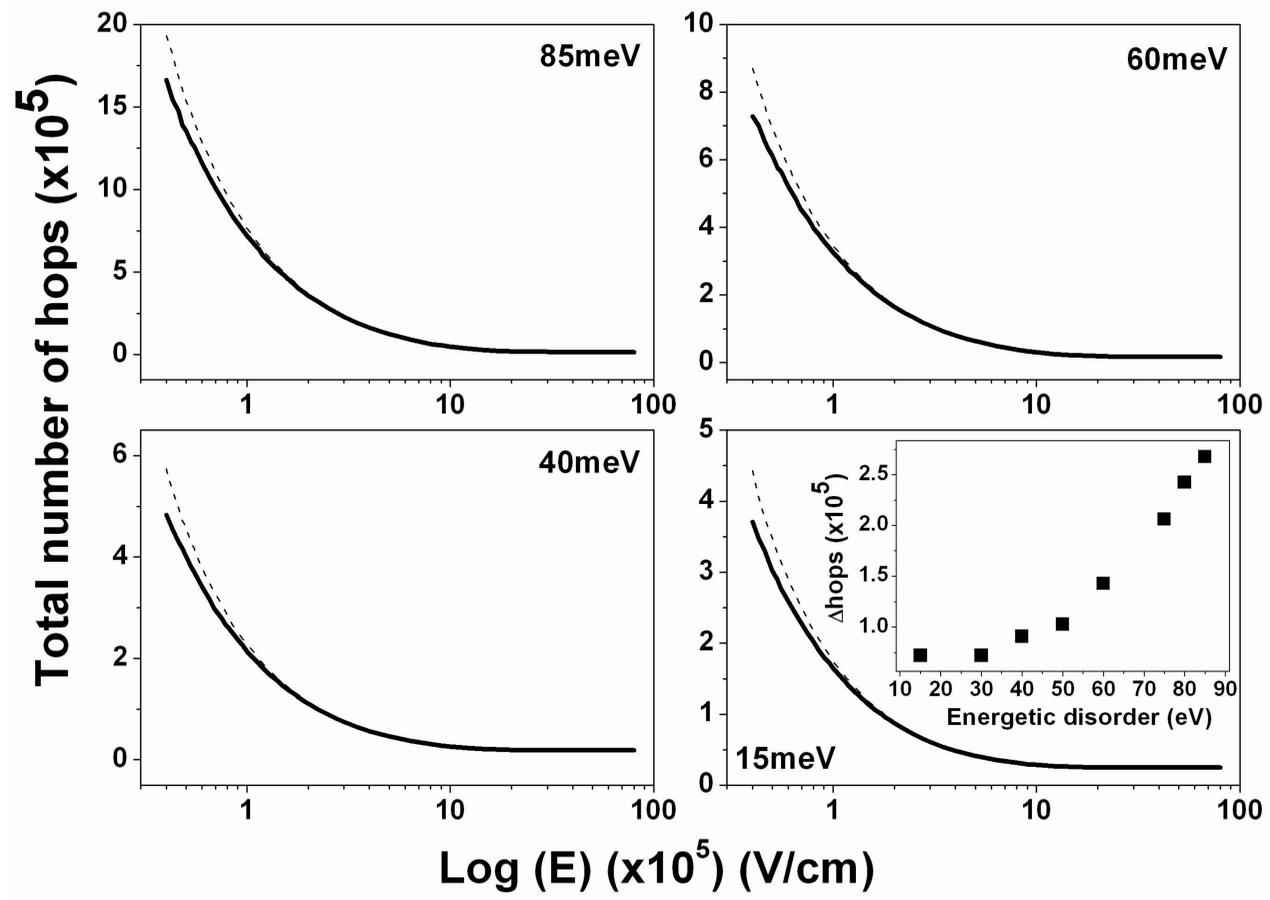

**Figure 7**

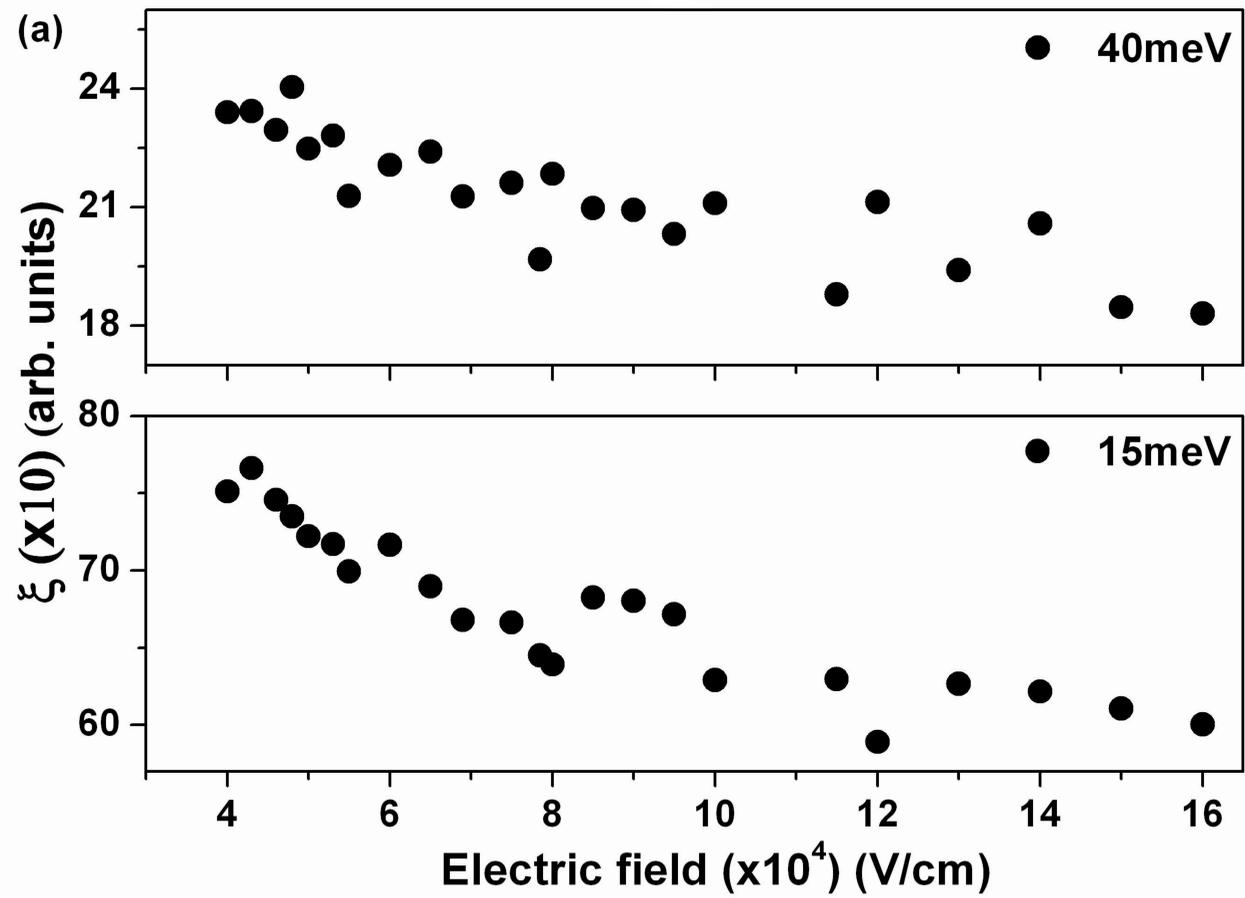

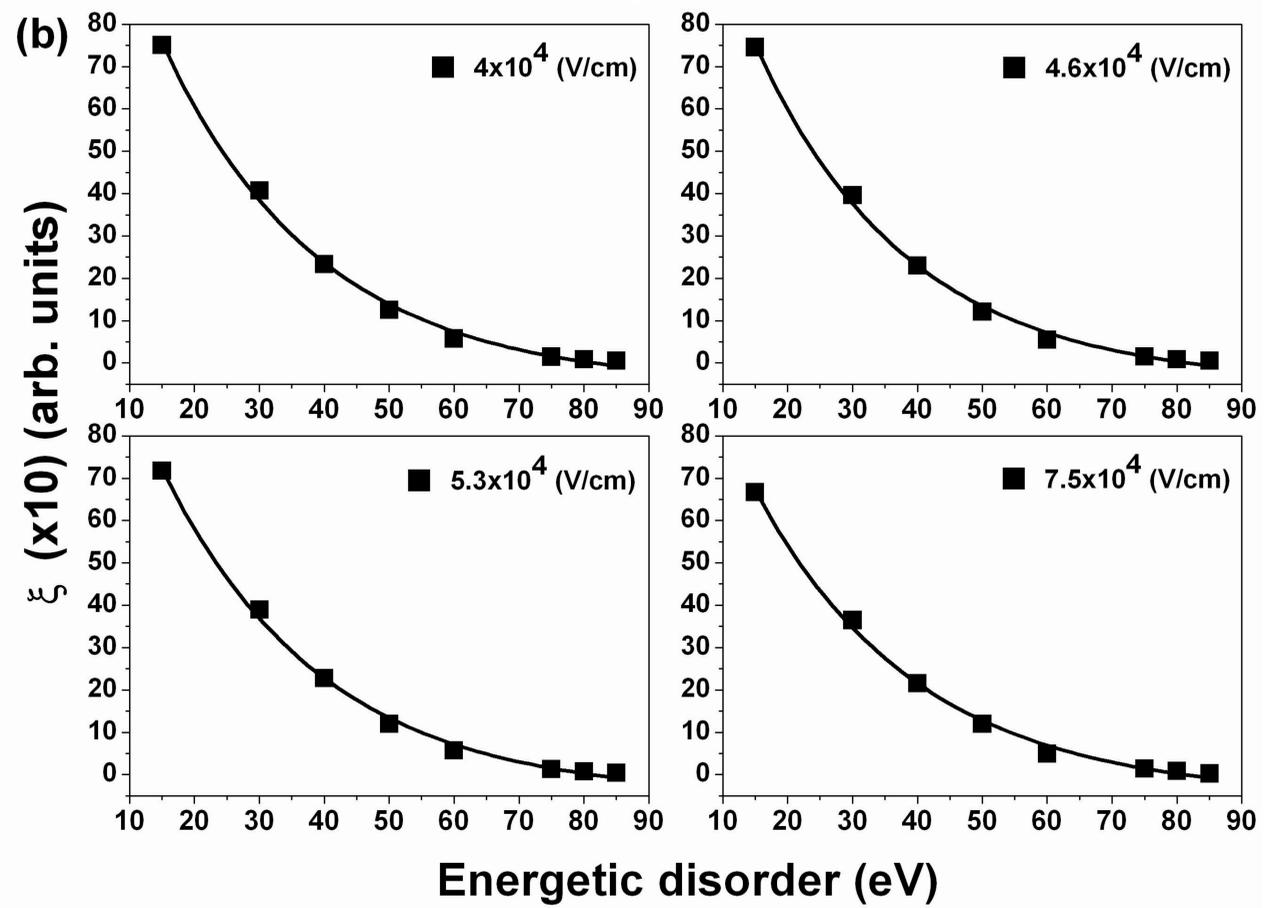